\documentclass[manuscript]{aastex}

\usepackage{graphicx}  

\def\kms{km s$^{-1}$}

\shorttitle{The First Spectroscopically Resolved Sub-parsec Orbit of a SBBH}
           
\shortauthors{Bon et al.}

\begin{document}

\title{The First Spectroscopically Resolved Sub-parsec Orbit of a Supermassive 
Binary Black Hole}

\author{E. Bon\altaffilmark{1,2}, P. Jovanovi\'{c}\altaffilmark{1,2}, P.
Marziani\altaffilmark{3,4},  A. I. Shapovalova\altaffilmark{5}, N.
Bon\altaffilmark{1,2}, V. Borka Jovanovi\'{c}\altaffilmark{6,2},  D.
Borka\altaffilmark{6,2}, J. Sulentic\altaffilmark{4} \& L. \v C.
Popovi\'c\altaffilmark{1,2}}

\altaffiltext{1}{Astronomical Observatory, Volgina 7, 11060 Belgrade, Serbia;}
\altaffiltext{2}{Isaac Newton Institute of Chile, Yugoslavia Branch;
Belgrde, Serbia;}
\altaffiltext{3}{INAF, Osservatorio Astronomico di Padova, Padova, Italy;}
\altaffiltext{4}{Instituto de Astrof\'isica de Andaluc\'ia, CSIC, Apdo. 3004,
18080, Granada, Spain;}
\altaffiltext{5}{Special Astrophysical Observatory of the Russian AS, Nizhnij
Arkhyz, Karachaevo-Cherkesia 369167, Russia;}
\altaffiltext{6}{Atomic Physics Laboratory (040), Vin\v{c}a Institute of Nuclear
Sciences, University of Belgrade, P.O. Box 522, 11001 Belgrade, Serbia.}

\begin{abstract}

One of the most intriguing scenarios proposed to explain how active galactic nuclei are triggered
involves the existence of a supermassive binary black hole system 
in their cores. Here we present an observational evidence for the first
spectroscopically resolved sub-parsec orbit of a such system in the core of
Seyfert galaxy NGC 4151.
Using a method  similar to those typically applied for spectroscopic binary
stars we obtained radial velocity curves of the supermassive binary system, from
which we calculated orbital elements and made estimates about the masses of
components. 
Our analysis shows that periodic variations in the light and radial velocity
curves can be accounted for an eccentric, sub-parsec Keplerian orbit of a
15.9-year period. The flux maximum in the lightcurve correspond to the
approaching phase of a secondary component towards the observer. 
According to the obtained results we speculate that the periodic variations in
the observed H$\alpha$\ line  shape and flux are due to shock waves generated by
the supersonic motion of the components through the surrounding medium. 
Given the large observational effort needed to reveal this spectroscopically
resolved binary orbital motion we suggest that many such systems may exist in
similar objects even if they are hard to find. Detecting more of them will
provide us with insight into black hole mass growth process.

\end{abstract}

\keywords{galaxies: active --- galaxies: interactions --- (galaxies:) quasars: individual (NGC 4151) --- shock waves ---
galaxies: Seyfert ---  black hole physics }

\section{Introduction}
Different mechanisms have been proposed to explain BLR variability in
AGN \citep[see for e.g.][and the references therein]{Gask03,Gask09}.
One of the most interesting possibilities involves the existence of a binary
system in the core \citep[see for e.g.][and the references 
therein]{Komossa,Bog08,Gask09,Ts2011,erac2012,Pop2012}. 
If black hole (BH) mass grows via major mergers we might expect to see the signature of a binary black 
hole in some or many active galaxies, if the merger process involves slow 
coalescence of the two components. 

NGC 4151, one of the best studied Seyfert galaxies, shows a complex sub-parsec structure. 
 Studies at radio,  visible, and X-ray wavelengths indicate violent processes including 
ejection of gas. Detection of outflow scales ranging from several hundred parsecs down to sub-parsec scales 
reveal collimated jets and biconical outflows  \citep[see e.g.][and the
references therein]{krae01,mund03,krae08,mund99}.
Violent variability in this active galactic nucleus (AGN) has been monitored over many years 
 \citep{shap08}. Earlier work showed variation in the wings of the CIV line  \citep{ulri85} 
that were associated to two localized emitting regions with line of sight 
velocities of -6100 and +8500 \kms\ with respect to the systemic velocity. These localized regions were thought to 
be the possible signatures of orbiting matter or of a two-sided jet.  A radio
jet or highly inclined disk (length $\sim$ 1 pc) aligned perpendicular to the
arcsec-scale radio jet was imaged at 6 cm  \citep{ulve98}.  
Flux variations in the optical domain have been monitored over a much longer time (over 100 
years)  \citep[see for e.g.][and the references therein]{Guo06,Okny07}. Optical
and UV emission is unresolved so clues about central structure come mainly from
analysis of broad emission line profile. Reverberation mapping 
studies  \citep[see for e.g.][and the references therein]{Onken07,shap08}
suggest a very compact broad line region (BLR) with broad Balmer lines
responding to continuum changes within a few days.
Many authors claim to find long term periodicities in the flux variations of NGC 4151
 \citep{Longo96} of e.g. 15.8 years 
\citep{Okny78,Okny07,Guo06,Pach83,Chuv08}. 
Compared to Seyfert 1s, NGC 4151 has a harder X-ray spectrum and  also similar to
black-hole binaries in the hard state \citep{Lubinski10}.

In this paper we analyze the H$\alpha$\ line shape and flux
variability in NGC 4151 during period of more than 20 years in order 
to investigate the BLR structure in NGC 4151. The possibility of periodicity in the line 
profile variations leads us to interpret it in terms of binary system
orbital motion at sub-parsec scale in the center of NGC 4151. 


The paper is organized as follows: in Section 2 we briefly
describe our data set, method, and analysis used to detect
periodicity in both the line flux and the shape of the profiles.
In Section 3, we discus possible scenarios for the obtained
periodic variations, and give evidence of a sub-parsec
supermassive binary BH system in the core of NGC 4151. In
Section 4, we discuss alternative models, and implications
of our results. Finally, in Section 5, we summarize
results and give conclusions.

\section{Analysis}

We analyzed 115 spectra of the H$\alpha$\ line profile in NGC 4151 covering more 
than 20 years. Most of the spectra were obtained during a period of 11 years
from 1996 to 2006 \citep{shap08,shap10}. This dataset was supplemented with: (i)
AGN watch spectra observed for three months beginning  1993 November 14
\citep{Kaspi96} with the 1m Wise Telescope \footnote{The AGN watch data
could be obtained in the digital format from the following link:
http://www.astronomy.ohio-state.edu/~agnwatch/data.html},
{
(ii) one spectrum from Asiago 1.82 m Ekar telescope of the Padova Astronomical
Observatory (1989 March 24),
which was equipped with a Boller \&
Chivens spectrograph and a 600 gr/mm grating yielded a resolution of 4--5 \AA\
FWHM, with PA=90, at the Cassegrain focus
and 
(iii) a spectrum from 1986 March 29 \citep{Ho95}\footnote{The spectrum from
Ho et
al. 1995 is
public and could be  obtained in digital format from the NED:
{http://ned.ipac.caltech.edu}}. }
All spectra were reduced to a common resolution 
of 15 \AA\ (about $\sigma = $ 340 \kms). The continuum was removed by subtracting a 
linear fit between 6265 and 6830 \AA. All spectra were scaled to the flux of [OI] 6300 \AA\
which is assumed constant at all epochs.

\subsection{Flux Variations}

We derived H$\alpha$ light curves  for the total flux (see Fig. 1) 
as well as for the partial flux in  1000 \kms  velocity bins (Fig. 2). 
We analyzed the light curves using Lomb-Scargle spectral analysis \citep{lomb76,scar82} procedures.

The Lomb-Scargle normalized periodogram \citep{lomb76,scar82,pres96} is a powerful tool for finding 
and testing the significance of weak periodic signals embedded in otherwise random and 
unevenly sampled observations. The significance level (i.e. the probability that the data contain 
a periodicity) can be tested fairly rigorously by the statistic $1 - p$, where $p$ is the so-called 
``false alarm probability''. This estimates the probability that the peak occurred in the presence of
pure independently and normally (Gaussian) distributed noise. A small value for
the false-alarm probability indicates a significant periodic signal. Resultant
periodicities were subsequently tested by fitting with sine functions (see
Fig. 1, bottom panel).

Since, we detected the periodicity in the H$\alpha$ line lightcurve (see Fig. 1), 
we tried to identify the parts of the line profiles that are most affected by
those periodical variations. For that purpose we analyzed the light curves  
for each 1000 \kms\ radial velocity bin of H$\alpha$ spectra, with significance
level above 90 \% (see Fig. 2). Since, we detected periodicity in certain light
curves (mainly in the core of the line for the velocity bins between -3000 and
4000 \kms), we tried to uncover possible mechanisms that could produce such flux
variations. For that purpose we tested many different line decomposition models,
taking into account the parts of the line where we detected significant
periodicity.

\subsection{Radial Velocity Curves}

We used Gaussian analysis to measure profile shape variations. This was
accomplished by fitting each H$\alpha$ profile with the sum of Gaussians
assuming some constrains.
The intensity ratio of narrow components (H$\alpha$ line, [OI]6300,6364\ \AA, [NII]6548,6584\ \AA, 
[SII]6717,6731\ \AA\ and HeI 6678\ \AA\ ) were modeled using spectra obtained during 
the minimum activity stage (2005 May 12). They were fitted with sum of
Gaussian profiles assuming that the intensity ratios of the narrow lines did not
change during the monitoring period. The widths of the narrow lines were assumed
equal during each fit.

The broad H$\alpha$ profile  was fit together with the narrow lines. We tested a number 
of models including those with more than one Gaussian component \citep[see for
e.g.][]{Shen10} and analyzed the light  curves of the component fluxes as well
as the radial velocity curves of each component during the monitoring period.
The choice of component widths was not arbitrary. We started from the
decomposition proposed in \citet{Sulentic00,Marz10}.
We noticed that a two component model \citep{Bon09,Bon06} consisting of a very
broad (covering the far wings) and broad (covering the line core) components
could not properly explain an intermediate width feature visible on the
H$\alpha$ wing ($\sigma = $ 600 \kms).
R and VBLR components are typical of Population B AGN \citep{Sulentic00,Marz10}
while displaced subpeaks have been occasionally observed in this class \citep{Zamfir10}.
The intermediate width subpeak was observed on both the blue and red wings of H$\alpha$ at 
different epochs but never at the same time leading us to interpret it as a single emitting 
component ($\sigma = $ 600 \kms). 

We then tried many fits with different widths for each of the three H$\alpha$ components  and 
adopted a model that consists of: (i) a very broad component (VBC) corresponding
to the far wings ($\sigma$ = 3400 \kms), (ii) a central broad component (CBC)
corresponding to the ``classical'' 
broad core of the line ($\sigma$ = 1700 \kms) and (iii) an intermediate width
component (bump) appearing on either the blue or red  wing of the H$\alpha$
profile
$\sigma$ = 600 \kms). The choice of Gaussian component widths were made in such
a way that each component width differs for at least two times, providing a
clear identification of each component. We fit the H$\alpha$\ profile using
three components with fixed widths. The CBC shift was fixed to the shift of the
narrow lines.

{To secure the identification of the Gaussian components we adopted the
model with three Gaussians with several important limitations according to the
shapes of profiles. We measured  the width of the bump on profiles where it
could be clearly isolated from other components. We found that the width of the
bump's Gaussian is around $\sigma$=600 \kms. We also tested the widths of other
two components and found that the Gaussian suitable to cover the central part of
the line profiles (CBC) needed to be at least twice the width of the bump's
Gaussian.
Also, we fixed the shift of the CBC  since we noticed that  the  shift of this
component was showing very small variations (smaller than the estimated errors
for this parameter), since the very broad wings (which were significantly
building up  in many profiles) could be covered only with very broad Gaussian
(at least twice the width of the CBC). We noticed that when the bump component
is present on the red side of  the profile, it was appearing either as a
separated peak, or as feature that was making the red side of the CBC much
steeper
than the regular shape of the Gaussian component that fit the core of
the 
line (CBC). When the bump component was appearing on the blue side, it was
just changing the steepness of the blue side of the profile, making it much
steeper than the shape of the Gaussian of the CBC. In this way including
the bump component was justified in our model. Similarly, it was necessary to
introduce the VBC Gaussian to fit the  far wings in some
profiles, that also could not be reproduced by other two components. 

In the case where we vary the intensities of the CBC and the VBC, the position
of the bump would shift, but for the solutions where the fit looks reasonable,
the shifts of the bump are smaller than the estimated errors. For example, if
intensities are changed for less than 5\%, the shift of the bump is changed, but
stays inside the error interval, while reduced $\chi^2$ increases for more than
an order of magnitude. If the variations of intensities are even larger the fit
becomes unreasonable, and do not follow the shape of the profile. 
}

Examples of fits for two epochs are shown  in Fig. 3. We noticed that the CBC
{
and VBC give major contributions to the total flux of the
line, and hence their lightcurves follow the total H$\alpha$ line lightcurve. On
these lightcurves (see Fig. 4. bottom) one can identify one dominant peak in
the interval of the Modified Julian Date (MJD)
between 50000 and 51000 MJD and another smaller peak between 52500 and 53000
MJD. Since we found similar  behavior in the lightcurves of all components, as
well as in the continuum and the  H$\alpha$  line, we concluded that the flux
variations are real in all components, and are not the caused by of the Gaussian
fitting procedure. More spectra from each observational campaign with
corresponding fits are presented in Figures 5-10}
 
 {\cite{Bog08} made simulations of H$\alpha$ emission line profiles in the
case of a supermassive binary black hole system with masses of
$10^7$ and $10^8$ $M_\odot$ and the following orbital elements:
15.7 year period and semimajor axis of $a=0.01$ pc. It is not that
obvious but one can see in their Figure 18 that the emission line profiles show
bump peak that appears on the blue side of the profile for the first part of
the orbital
period, and in the second part of the orbital period the bump peak starts to
appear on the red side of the profile. Also, one can see that the central peak
is present in the most of profiles as well as the wider structure (VBC) even
though we approximated it with a Gaussian instead of more complex profile as
seen in
this figure (probably a disk like emission profile). In our case there was no
significant difference in the radial velocity curves and light curves in the
case of fitting using a VBC Gaussian instead of a VBC disk profile.}

We analyzed radial velocity variations in the VBC and bump components.
We also fit radial velocity curves assuming Keplerian 
orbits using the ``Velocity'' code \citep{Velocity}. The uncertainty in the
position of the bump 
was estimated to be $\pm$ 200 \kms\ (when detected on the blue side) and  $\pm$
70 \kms\ (on the red) at 2$\sigma$ confidence level ($P$=0.99). Estimated
uncertainties for radial velocity of the VBC were around $\pm$ 300 \kms. These
errors were set by $\chi^2$\ behavior when varying only the  
displacement from the best fit radial velocity and are regarded as minimum
uncertainties.

{
To test the influence of the fitting procedure on the lightcurves and radial
velocity curves we varied the widths of each component.
For the bump we
found that in some cases the fit becomes unreasonable if we change its width
by more than 100 \kms. For the change in the widths of the VBC and CBC,
the radial
velocity curves of the bump component did not change much (a few times less than
the estimated errors), even thought the VBC width was varied from 3000 to 4500
\kms,
and CBC from 1500 to 2000 \kms. We noticed that for the broader VBC, the
amplitude
of the radial velocity curve of that component, became larger, but kept the
same shape and behavior, and hence led to the similar orbital parameters.
However, in that case the obtained mass ratio between secondary (bump) and 
primary (VBC) will vary between 0.2 (for $\sigma_{\rm VBC}$ = 3000 \kms) and 0.6
(for $\sigma_{\rm VBC}$ = 4500 \kms).
Therefore, this method gives more reliable estimations of the orbital elements
than those for the mass ratio. 

We are aware that, when the bump is close to the
line core (at $v_r \sim$ 0) its position is well constrained because of the
model we build with three symmetric Gaussians, since the bump is overwhelmed by
the BC and VBC emission so as to become not visually identifiable. As the bump
radial velocity changes sign rather suddenly, this condition occurs only for a
few epochs.
}

\section{Results}

Light curve analysis  is presented in Figs. 1, and 2.
Fig. 1 shows that the total flux could be well fit with a sine function with
5700 $\pm$ 300 day periodicity. Fig. 2 shows Lomb-Scargle analysis fluxes for
each 1000 \kms\ radial velocity bin of H$\alpha$ with significance level above
90 \%. The resultant values ranged from $P\approx$ 5000 to 6000 days (see Fig.
2).

\subsection{Periodical Variations in Lightcurves and Possible Mechanisms}

The detection of a significant periodicity is a major result of this study. The
analysis of the light curves showed a periodicity of nearly 16 years, which is
in agreement with previous studies (of 16 year periodicity) obtained (using
different methods) for a much earlier monitoring period, covering few earlier
cycles \citep[see][]{Okny78}.

There are several possible sources  of periodicity.  
Apart from the orbital period ($T_\mathrm{P} \approx 5.5 r_{16}^{-\frac{3}{2}}
M_{7}^{-\frac{1}{2}} \mathrm{yr}$), relevant timescales include the timescale
associated with geodetic precession $T_{P} \approx 2 \cdot 10^{4}\-  
r_{16}^{-\frac{5}{2}} (\frac{M}{m}) M_{7}^{-\frac{3}{2}}~~\mathrm{yr}$ (i.e.,
precession of the BH spin around the total angular momentum of the system
\citep{Beg80}), and accretion disk precession due to a second massive black hole
 $T_{P,AD} \approx 23 \cdot r_{16}^{3}  (\frac{M}{m})^{\frac{1}{2}}
m_{6}^{-\frac{1}{2}} R_{d,16}^{-\frac{3}{2}} \cos^{-1} \theta_{0}~ \mathrm{yr}$,
where $R_{d}$ is the accretion disk radius \citep{Katz97}. Measured parameters
for NGC 4151 imply that the geodetic precession is $T_{P} \sim 10^{5} $ yr.  
Accretion disk precession is expected in any AGN hosting a binary black hole and
$T_{P,AD}$ is much lower than  the value estimated for geodetic precession.
However for NGC 4151  $T_{P,AD}\approx 250 R_{d,16}^{-\frac{3}{2}} \cos^{-1}
\theta_{0}$ is still 
significantly longer than the derived period. A third possibility is that
radiation force  induce a ``self-warping'' of the accretion disk (obviating the
necessity for a second black hole). In this case the period would be
\citep{Pring96} $T_\mathrm{P,warp} \approx 13.6  \cdot    ({M}_\mathrm{ion.
gas,-2}) M_{7}^{\frac{1}{2}} R_\mathrm{d,10^{3}R_\mathrm{g}}^{\frac{1}{2}}
{L}_{44}^{-1}~~ \mathrm{yr}$. The self-warping timescale is consistent with the
observed $P$ and could, in principle, reinforce a periodic behavior also
involving continuum fluxes. 

\subsection{A Sub-parsec Supermassive Black Hole Binary}

We analyzed radial velocity curves for the VBC and bump components, because we
noticed large variations  of their velocity shifts. Fig. 4 indicates that the 
radial velocity curve can be well described with Keplerian orbits in a binary
system. 

The derived orbit is eccentric ($e$ = 0.42) with a period of $P$ = 5776 days and
longitude of pericenter $\omega\approx 95^\circ$. We note that orbital
precession would be expected but we did not take it into  account since the
predicted $\delta\Phi < 1^\circ$ per period in Schwarzschild metric. Using
Kepler's laws we estimated the semimajor axes: $a_1 \sin i = 0.002$ pc, $a_2
\sin i = 0.008$ pc and masses:  $m_1  \sin^3 i =3 \cdot 10^7  M_\odot$ and $m_2
\sin^3 i = 8.5 \cdot 10^6  M_\odot$ (where $i$ is the inclination of the orbital
plane).
These results are in a good agreement with the corresponding theoretical results
obtained using numerical simulations \citep{Bog08} which showed simulations
(including the emission line profiles) of an eccentric Keplerian system with
15.7 year period, at 0.01 pc distance with similar total mass as found here. We
also, estimated semimajor axes and masses for different inclination angles and
found that an inclination of about $65^\circ$ yields semimajor axes  of 0.0024
pc and 0.0094 pc, while the masses of each component are $4.4 \cdot 10^7$
$M_\odot$and $1.2 \cdot 10^7$ $M_\odot$. Our mass estimate is in good agreement
with the values obtained from other methods e.g. reverberation mapping
\citep{Onken07}. The derived separation of two supermassive black holes (SMBHs)
of about 0.01 pc is expected from theoretical models for the evolution of binary
SMBHs \citep{MM2001}. The derived period is in agreement with previous studies
of optical variability in NGC 4151 \citep{Okny78,Okny07,Guo06}, which indicate
existence of previous cycles \citep[see][]{Okny78}.

If this model is correct then gravitational radiation should lead to
coalescence \citep{1973grav.book.....M} in  $\sim$ 10$^{8}$ yr. The evolution of
a BBH system through merging is a topic of much recent research
\citep[see][for an exhaustive review]{MM05}. Regardless of the  evolutionary
stages governed by stellar  dynamics, there will be a final  stage when 
dissipation of orbital energy is due to emission of gravitational radiation.
When the binary separation is on sub-parsec scales, coalescence time scales
$\propto M^{-3}$\ imply that such binary configurations could be relatively
long-lived in low-mass AGNs, if other mechanisms do not intervene
\citep[e.g.][]{Hayasaki08}. Systems like  NGC 4151 with a central supermassive
black hole and a (or perhaps, even more than one) much less massive satellite
could be common even if they are hard to find.

Considering the relatively large uncertainty in radial velocity for the VBC,
the mass ratio could be different than we have derived. In the most extreme case
we could be observing some {cloud of}
gas \citep{cloud} spiraling down toward the center of attraction, but then we
should not expect a very long lifetime for 
such cloud due to dissipation making this scenario less probable, since several
16 year cycles are already reported \citep[][including the result of this
paper]{Okny78,Okny07,Guo06}.
If the secondary object is compact but with a mass much smaller than the primary
component, then its orbit would be defined with the bump component only,
implying a distance of about 0.01 pc and the same period (about 15.9 years
and eccentricity of about 0.4).

\section{Discussion}

If we assume that a second massive black hole is indeed present,
then the pretty large value of eccentricity suggests an analogy with OJ 287
\citep{Valtonen12}, where its orbital motion is assumed to be an explanation of
its flux variation. Orbital motion of SMBHs has been recently reported in 3C 66B
\citep{Sudou03} and earlier in 3C390.3 \citep{Gask96}. We are however, far from
the extreme conditions proposed for OJ 287, a source with a black hole mass
estimated to be $\sim$500 times larger then the one of NGC 4151. As a
consequence, we expect a small pericenter precession. Energy loss due to
gravitational radiation is very small if compared to OJ 287.  Nevertheless, this
system is among the ones where further evolution is governed by gravitational
losses. Since these systems are expected to occur when the binary separation
roughly corresponds to the size of the BLR, extensive monitoring may reveal
other cases (we remark that the observational coverage of NGC 4151 is almost
unique).  

If the component separation in a binary SMBH system is sub-parsec: i.e. on the
same order or smaller than the expected radius of the BLR, we have justification
for a  CBC fixed to the rest frame as derived from the redshift of the narrow
lines.  This implies that the orbit of the binary SMBH system is most probably 
located within the BLR region. The estimated orbital velocities ($v=v_r \cdot
\sin^{-1} i \cdot \sin^{-1} \phi$;  where $i$ represents inclination and $\phi$
the orbital phase) are above 4000 \kms\ for the bump and around 1000 \kms\ for
VBC
for the quadrature phase (where ${\rm sin} \phi = 1$ and the radial component is
highest towards the observer). These are much larger than the typical speed of
sound (typically of order of tens to hundreds of \kms) expected for such
conditions \citep{Bog08,dopita95,Mayer07}. Therefore, one would expect them to
produce shock waves in the BLR \citep{Shi12,Arty96,Mayer07}. The epoch with
maximum total flux (see Fig. 1, bottom panel), in the MJD interval between
50000 and 51000, corresponds to the epoch when the bump
(secondary) component is approaching the observer with supersonic velocity,
producing shock waves in the direction of the observer (see Fig. 4), while a
smaller
flux maximum that follows afterwards (MJD between 52000 and 53000) corresponds
to the epoch when the approaching shock wave is produced by the VBC.  The epochs
where the fluxes are at minimum (Fig. 4) correspond to the phase when the shock
waves are propagating perpendicular to the line-of-sight direction (conjunction
phase).  This  is consistent with previous results \citep{shap08}: 
non-radiatively heated emission is needed to explain an excess of line emission
with respect to the case of  pure photoionization during the epoch of the flux
maximum (MJD between 50000 and 51000). Shocks  can provide  a suitable heating
mechanism. 

In other scenarios, involving self-warping and flares, one would expect
circular concentric orbits with sinusoidally shaped radial velocities. This
differs from the radial velocity curves in Fig. 4 that corresponds to the more
eccentric Keplerian orbits. Also, the simulated light and radial velocity curves
of an
elliptical disk, single spiral arm or fragmented spiral arm presented in
\citet{lewi10}, could not describe well the shape of the light and radial
velocity curves observed in NGC4151.

{One of the alternative explanations for peaks or shoulders
seen in the broad H$\alpha$ profiles could be the emission of an accretion disk.
To test such possibility we tried to fit the \cite{ch} model of accretion disk
into broad line profiles from different epochs, in such a way that the peak
of the bump would be fitted with one of two peaks from the disk profile. We
found that it could be possible to fit an accretion disk model in most of the
profiles, but always with needing the additional components. For example,
we could fit the disk profile into the core of the line in such way
that the bump component would be covered with one of the peaks from the disk.
In that case, the result for the inclination parameter is very small (around 15
degrees), which is not that likely explanation due to much higher inclination of
jets, around 60-70 degrees \citep[see][]{Das2005,krae01}. The variations in
observed profiles could be explained assuming that the emission lines originate
in  the precessing accretion-ring, or precessing elliptical accretion-ring
\citep{Storchi-Bergmann97}, with the variations of the width of the inner radius
of emitting ring, or with the shift of the disk profile (so the moving bump
would be fitted with one of peaks from the disk profile). In all these cases
it was necessary to involve additional components (e.g. VBC and
CBC). If the wider disk profiles were used to fit into the far wings, then
the corresponding inclination would be around 45 degrees, covering the far wings
of the
profiles, as in the VBC in our model, resulting in a similar behavior to the
VBC. This will include the need for the CBC and a bump component and with
radial velocity curves similar to those in our simplified model with three
Gaussians.
We found cases with a higher red peak  than blue, that could not be
explained
with the circular disk model, but could be explained with elliptical disk model
\citep{Eracleous95}. For this alternative model, its precession could
explain its rising and moving peaks. All alternative explanations
must be consistent with the obtained periodicity (see \textbf{\S 3.1}). Some
other
alternative explanations could be: orbital motion of a flare or a bright
spot \citep[see for e. g.][]{Newman97}, self-warping accretion disk
\citep{Pring96}, precession of an eccentric circular and elliptical accretion
disk, circular disk with spiral arm \citep[see for e. g.][]{lewi10}, the
precession of a bipolar outflow \citep[see for e. g.][]{CrenshawKramer2007},
and an
inhomogeneous disk \citep[see for e. g.][]{Eracleous95}. For the most of these
scenarios we should expect more sinusoidal-like behavior of radial velocity
curves \citep[see for e. g.][]{Newman97,Storchi-Bergmann97,lewi10}, which
differs from Fig. 4. For these alternative models, more tests should be
made, to prove or disprove any of them. 

The radial velocity curve is immediately consistent with orbital motion as we
are able to fit it with an eccentric orbit.  The idea of orbital motion seems
the most plausible. It is a consequence of the favorable value of the argument
of the periastron (i.e., a fortunate coincidence) that we are able to see a
sudden
change in the radial velocity. 
It is however what is expected and observed in many binaries with highly
eccentric orbits.}  

\section{Conclusion}

We have analyzed the shape and flux variations in the H$\alpha$ light curve
for the Seyfert galaxy  NGC 4151 using observations spanning 20 years. We also
studied variations in different  parts of the line profile. Lomb-Scargle
spectral analysis of these light curves suggest variations that can be
satisfactorily fit with sine functions and indicating periodicity  of about 5700
days for the total flux variation and around 5800 days for most of the 1000 
\kms\
H$\alpha$ intervals. Analysis of radial velocity curves for H$\alpha$ emission
line  components (VBC and bump) reveal evidence for orbital motion consistent
with the signature of a sub-parsec scale SMBH system with orbital period of
about 5780 days.

In this scenario periodic variations in the optical spectra of NGC 4151 could
be generated by supersonic motions of the components through an extended broad
line region. Maxima in the flux variations correspond to phase when a shock wave
is generated toward the observer. This mechanism is also able to explain
the detected
outbursts in the optical fluxes of emission lines and the continuum and is in
agreement with reported shock heating during the period around the light curve
maximum \citep[see][]{shap08}, as well as the similarity of its X-ray spectra to
black-hole binary spectra in the hard state \citep{Lubinski10}.

This interpretation is consistent with some previous theoretical results
\citep{Bog08,Mayer07,Mayer10,MM2001} and would support the idea of BH mass
growth by major merger rather than by slower accretion processes.
If the phenomenology outlined by our analysis is indeed general for all similar
AGN then an important question arises: Is binarity a necessary condition for AGN
ignition?
Support for such an assumption could be related to changes in spectral type
(Sy1--Sy1.5-Sy2)
manifested by NGC 4151. The activity type of the galaxy changes from Sy1 to
Sy1.5 when it is most active  \citep{shap08}
and  to Sy2 at its deep minimum phase, as it was in 1984
\citep{Penston84,Lyutyj}.
Different spectral types may be possible only at a certain phase of orbital
motion in the binary SMBH.
This opens the question whether the different activity types of active galaxies
correspond to different orbital phases of such systems and whether the binarity
is a necessary condition for the activity switch-on.
Further monitoring is needed to assess the persistence (or disappearance) of the
orbiting components. 
If our results would be confirmed by future monitoring campaigns or analysis on
different sets of data, then NGC 4151 might become an important candidate for
gravitational wave detection in the future.

 \acknowledgements
{
This research is part of projects 176003 ''Gravitation
and the large scale structure of the Universe'' and 176001 ''Astrophysical
spectroscopy of extragalactic objects'' supported by the Ministry of Education
and Science of the Republic of Serbia, contract P08-FQM-4205 from La Junta de
Andalucia and also grants N09-02-01136a and 12-02-01237a supported by RFBR. We
also thank Samir Salim and Giovanni La Mura for
constructive discussions.
}

\begin{figure}
\centering
\includegraphics[height=\columnwidth]{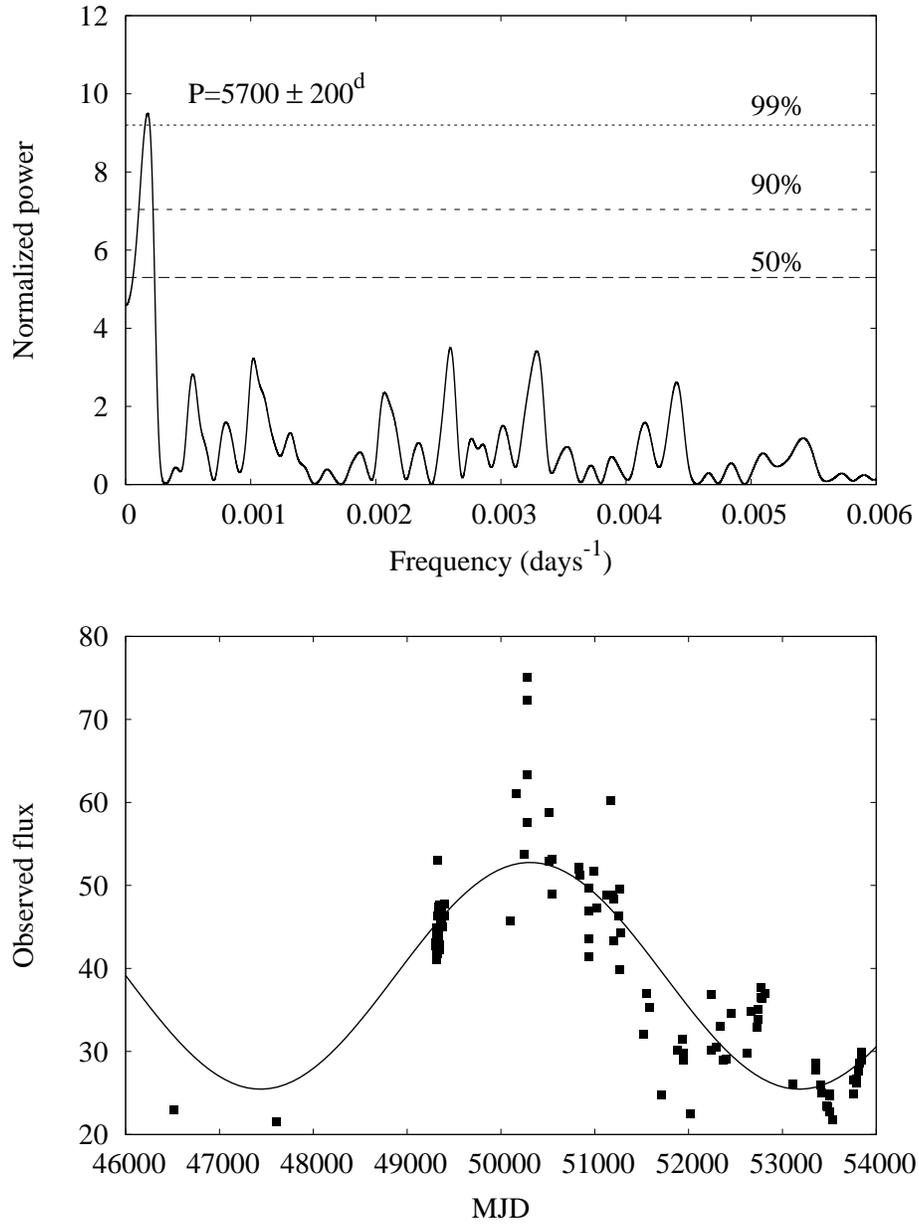}

\caption{
Lomb-Scargle normalized periodograms  of the
light curves for total flux (top) and the corresponding sinusoidal fit
(bottom).}
\label{lomb}
\end{figure}

\begin{figure}
\centering

\includegraphics[height=\columnwidth, angle=270]{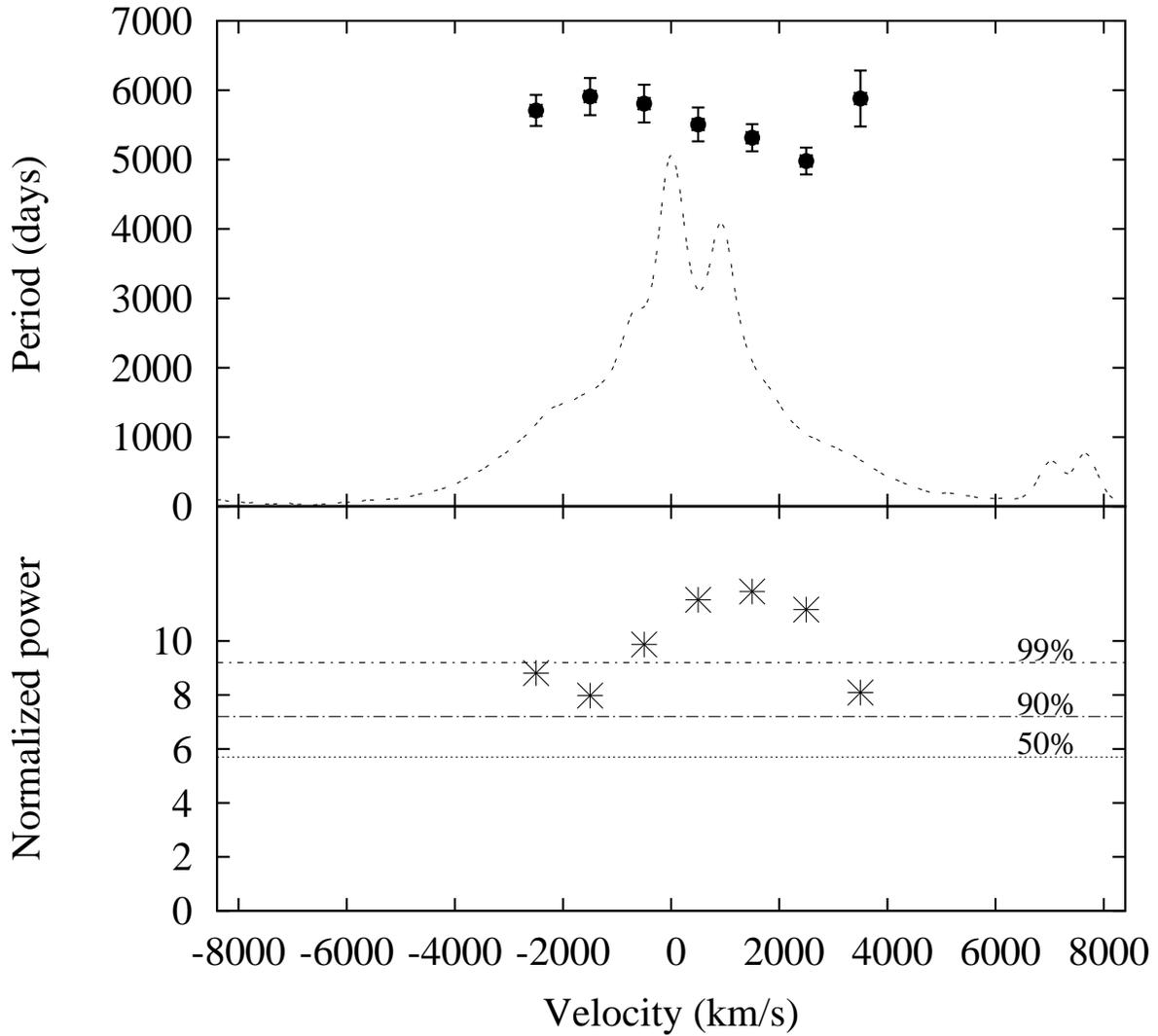}
\caption{
Period (filled circles with error bars) obtained for each 1000 \kms\ light
curve of H$\alpha$ line profile and their normalized powers (asterisks).
Horizontal straight lines represents 50\%, 90\% and 99\% normalized power
significance levels.}
\label{lombbin}
\end{figure}

\begin{figure}
\centering
\includegraphics[scale=.4, angle=270]{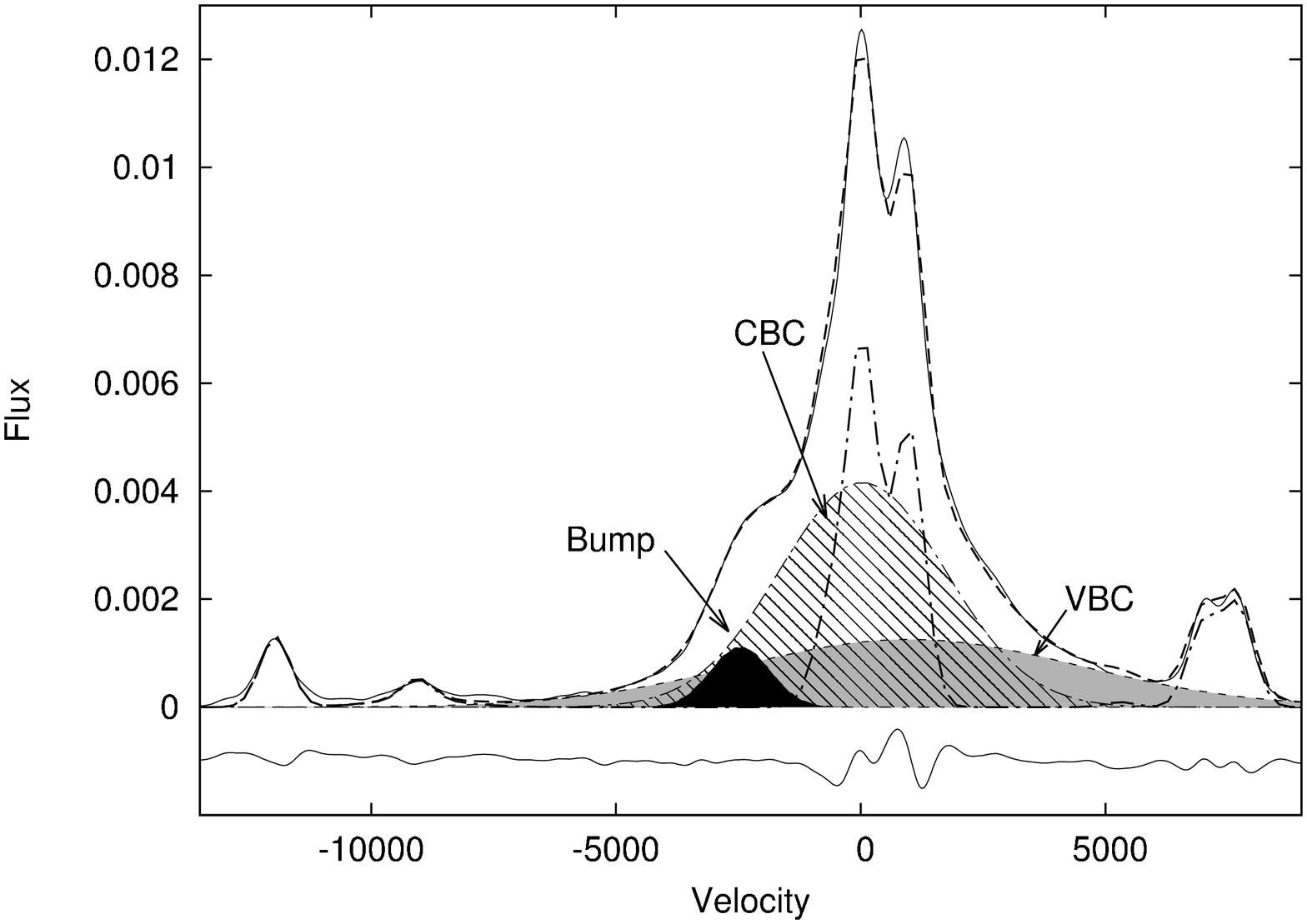}
\includegraphics[scale=.4, angle=270]{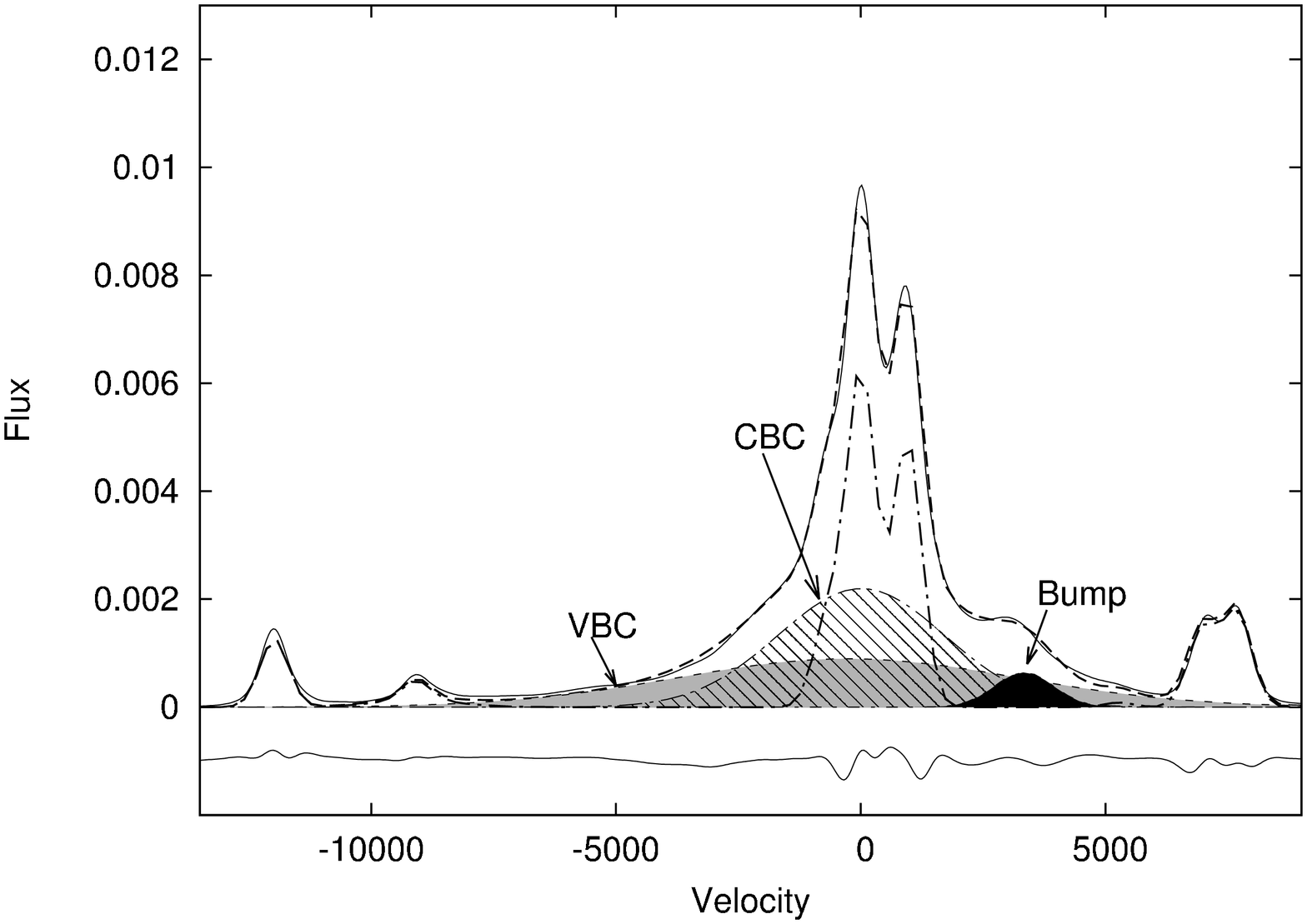}

\caption{
Examples for the best fit (dashed line) of the H$\alpha$ line profile (solid line) for two epochs: MJD 51203 (up)
and 52621 (down). Three Gaussian components: VBC ($\sigma$ = 3400 \kms, shaded
with gray), CBC
($\sigma$ = 1700 \kms, shaded with line pattern) and the bump ($\sigma$ = 600 \kms, shaded with black) component, are fitted
in the broad profile. The dash-dotted line represents the narrow
line template.
The panel at the bottom of the plots represents residuals of the fit.}
\label{fitexample}
\end{figure}

\begin{figure}
\centering
\includegraphics[scale=.5]{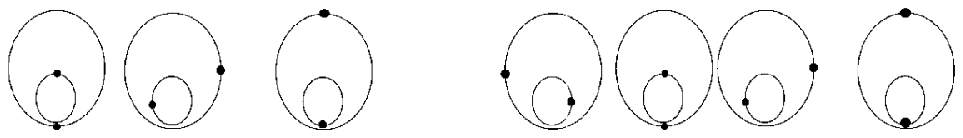}

\includegraphics[scale=.5]{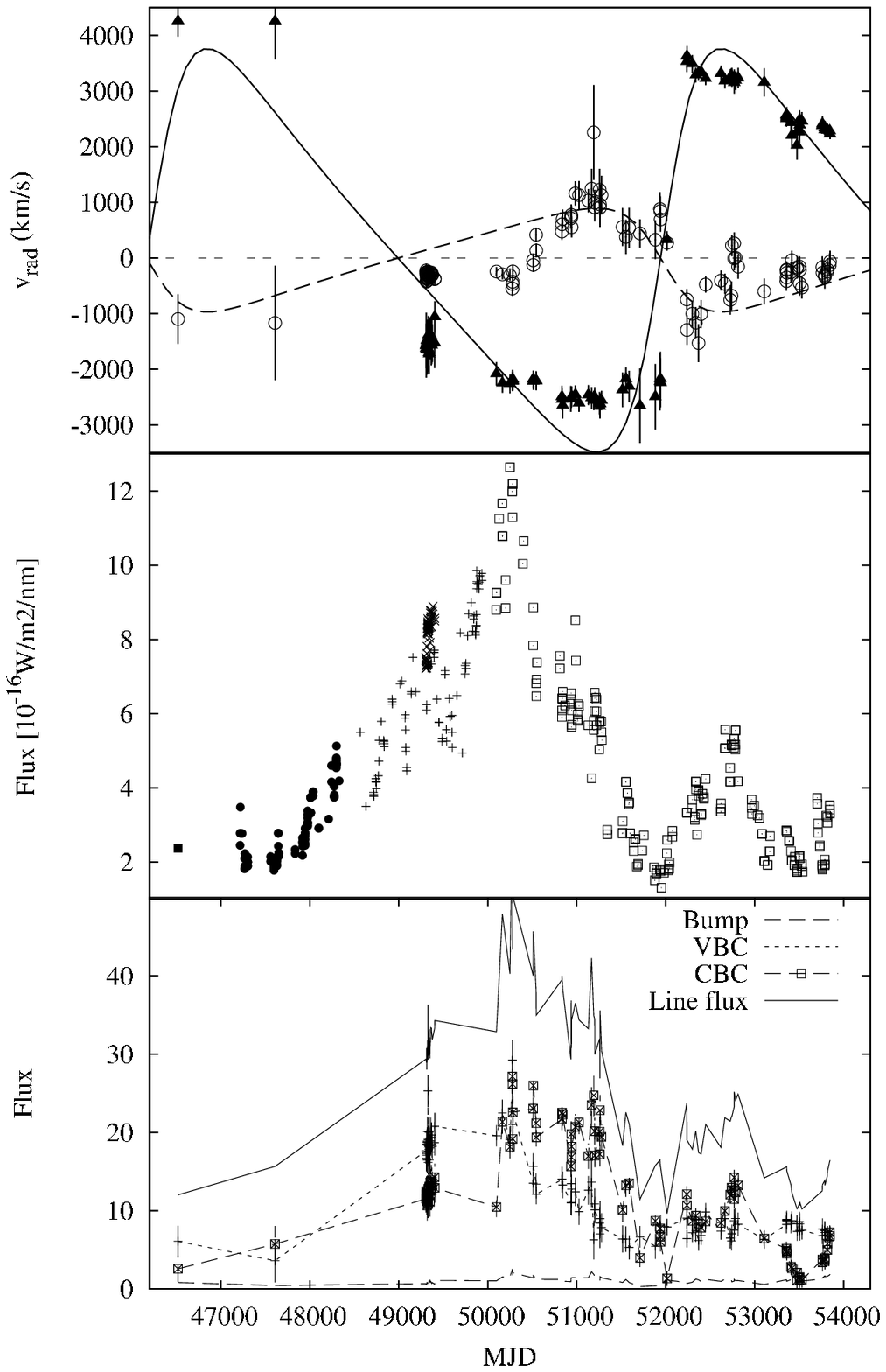}
\caption
{{ Keplerian orbits of a supermassive black hole binary.} 
Top: radial velocity curves of the bump (open triangles with y error bars)
and VBC (open circles with y error bars) obtained from Gaussian decomposition of
the broad H$\alpha$ line, as well as fitted radial velocity curves for orbits of
both components (solid and dashed lines). Middle: continuum flux light curves
for NGC4151 at  512.5 nm, compiled from \citet{Malkov97}  (plusses),
\citet{Kaspi96} (crosses), \citet{shap08} (open
squares), as well as at 656.3 nm from \citet{Sergeev94}  (filled circles). Bottom:
light curves of the relative fluxes (normalized to the flux of
OI line) with y errorbars of VBC (doted line), CBC (dashed line with open squares), bump
component (dashed line) and the H$\alpha$ line (solid line).
Illustration of the corresponding orbital phases are presented above
the radial velocity curve panel, assuming counter clockwise direction of each
black hole motion and the direction toward the observer below the plots.}
\label{radvel}
\end{figure}



\begin{figure}
\centering
\includegraphics[scale=.45, angle=270]{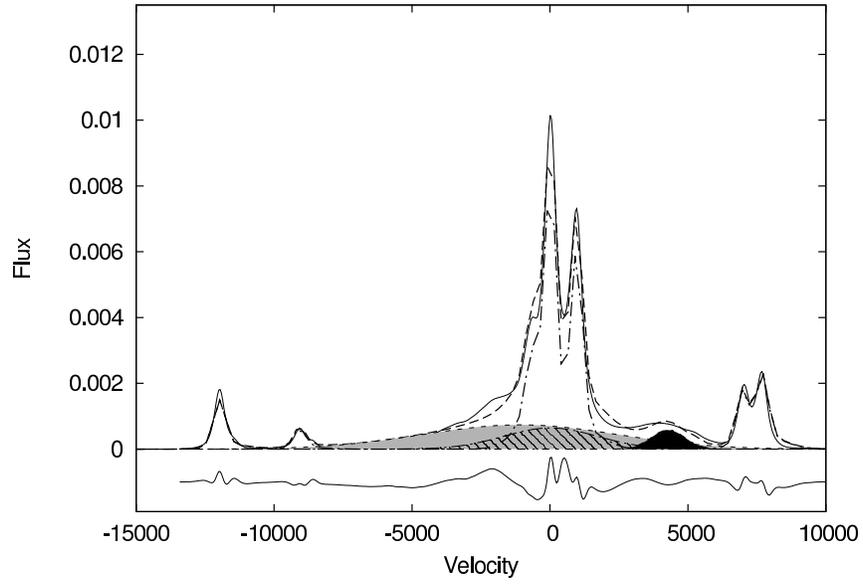}
\caption{
The same as in Fig. \ref{fitexample}
but for the spectrum from \citet{Ho95} (epoch MJD 46518.5).  
}
\label{fit2}
\end{figure}

\begin{figure}
\centering
\includegraphics[scale=.45, angle=270]{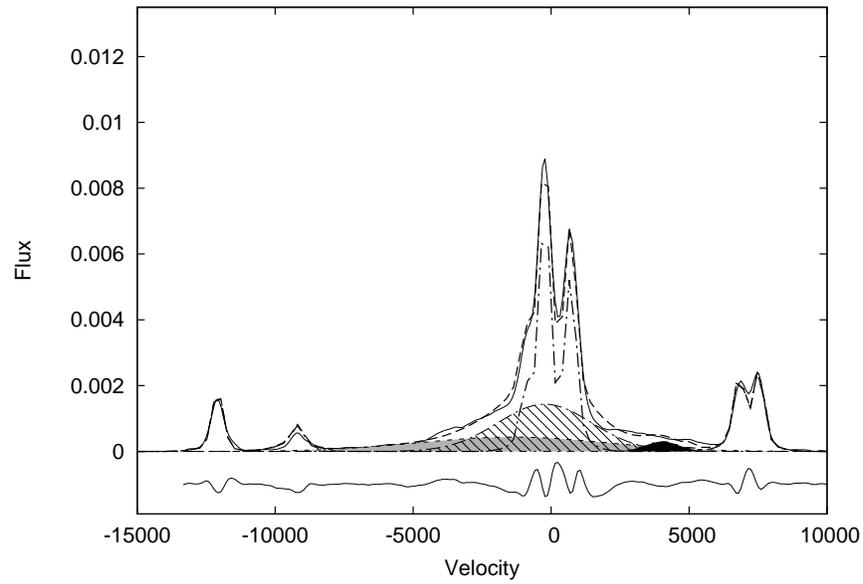}
\caption{The same as in Fig. \ref{fitexample},
but for the spectrum obtained
on Asiago 1.8 m Ekar telescope (epoch MJD 47609.529). 
}
\label{fit3}
\end{figure}

\begin{figure}
\centering
\includegraphics[scale=.45, angle=270]{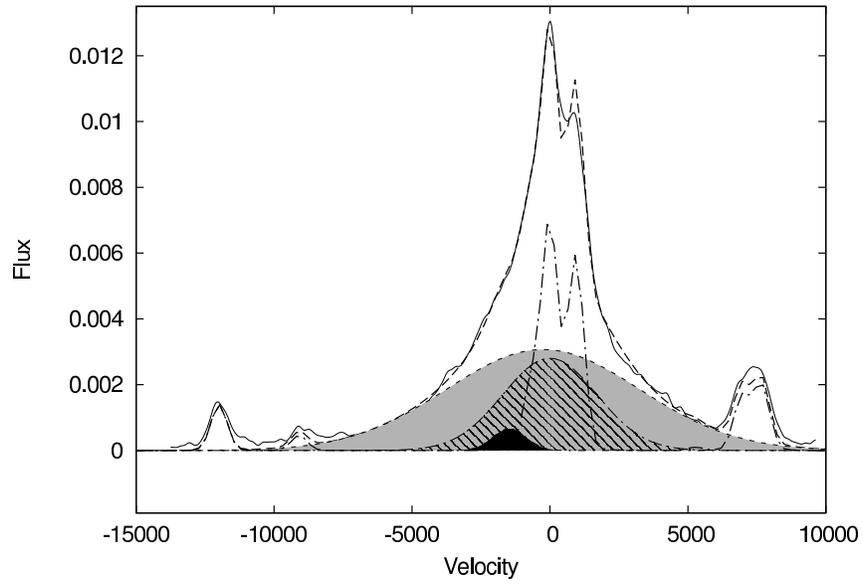}
\caption{The same as in Fig. \ref{fitexample}
but for the spectrum obtained
during AGN watch program \citep{Kaspi96} (epoch MJD 49327.577).
}
\label{fit4}
\end{figure}

\begin{figure}
\centering
\includegraphics[scale=.45, angle=270]{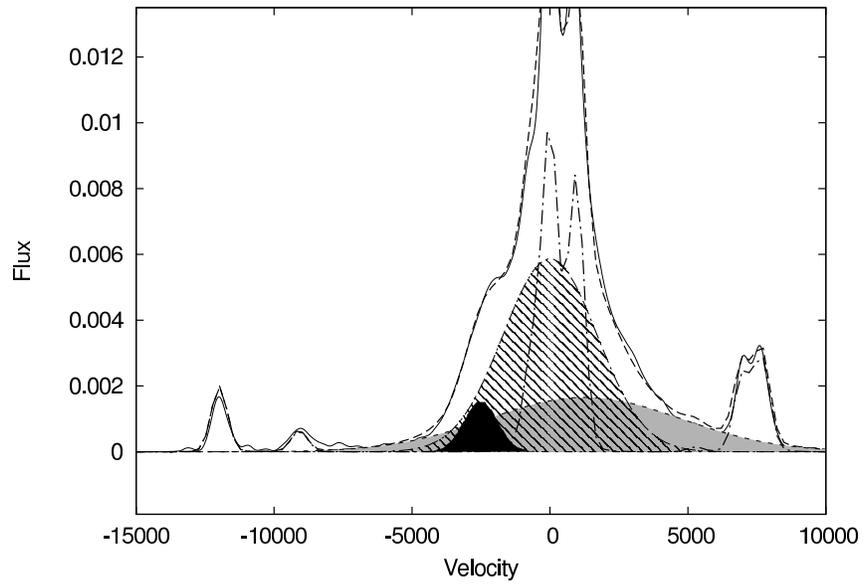}
\caption{The same as in Fig. \ref{fitexample}
but for the spectrum from the set of \citet{shap08,shap10} (epoch MJD 51166.656) 
with the bump component positioned in the blue
part of the spectrum.}
\label{fit5}
\end{figure}

\begin{figure}
\centering
\includegraphics[scale=.45, angle=270]{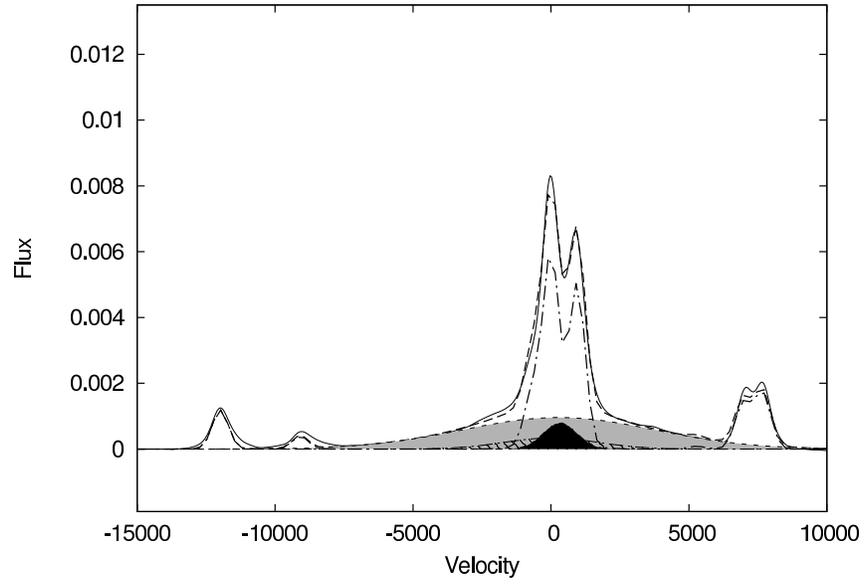}
\caption{The same as for Fig. \ref{fit5} but for the epoch MJD
52016.36, with the bump component positioned in the central
part of the spectrum.}
\label{fit6}
\end{figure}

\begin{figure}
\centering
\includegraphics[scale=.45, angle=270]{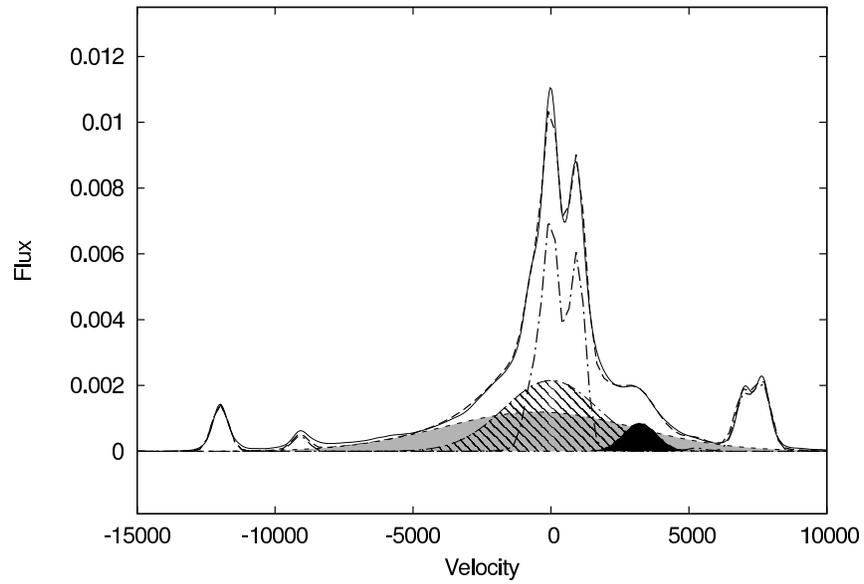}
\caption{The same as for Fig. \ref{fit5} but for the epoch MJD 52621, with
the bump component
positioned in the red
part of the spectrum.}
\label{fit7}
\end{figure}


\begin{thebibliography}{99}

\bibitem[Artymowicz \& Lubow (1996)] {Arty96} Artymowicz P. \& Lubow, S. H., 
\apj, 1996, 467, 2, L77.


\bibitem[Begelman et al. (1980)]{Beg80}  Begelman, M.~C.,
Blandford, R.~D., \& Rees, M.~J.\, 1980,  {\nat},  287, 307.

\bibitem[Bogdanovi\'c et al. (2008)]{Bog08} Bogdanovi\'c, T., S., Britton D.,
Sigurdsson, S., Eracleous, M., 
2008, \apjs {174} , 455.

\bibitem[Bon et al. (2009)]{Bon09} Bon, E., Popovi\'c, L. C\v., Gavrilovi\'c,
N., La Mura, G. , Mediavilla, E., 2009, {\mnras}, 400, 924.

\bibitem[Bon et al. (2006)]{Bon06} Bon, E.,Popovi\'c, L. \v C., Ili\'c, D.,
Mediavilla, E.,2006, \nar, 50, 716


\bibitem[Crenshaw and Kramer (2007)]{CrenshawKramer2007} Crenshaw and
Kramer 2007,\apj, 659,250.

\bibitem[Chuvaev et al. (2008)]{Chuv08} Chuvaev, K., K., Oknyanskij, V., L.,
Lyuty, V., M., 2008, {Izv.Kr.Astrof.Obs.}, 104, 109.

\bibitem[Chen \& Halpern (1989)]{ch} Chen, K., \& Halpern, J., P., 1989, ApJ,
344, 115

\bibitem[Das et al. (2005)] {Das2005} Das, V., Crenshaw, D. M., Hutchings, J.
B.; Deo, R. P., Kraemer, S. B., Gull, T. R., Kaiser, M. E., Nelson, C.
H., Weistrop, D., 2005, {AJ}, 130, 945.

\bibitem[Dopita M. A. (1995)]{dopita95} Dopita, M. A.,
1995, {ApSS}, {233}, 215.


\bibitem[Eracleous et al. (1995)] {Eracleous95} Eracleous, M., Livio, M.,
Halpern, J., P., Storchi-Bergmann, T., 1995, {ApJ}, 438, 610.

\bibitem[Eracleous et al. (2012)] {erac2012} Eracleous, M., Boroson, T., A.,
Halpern, J., P., Liu, J., 2012, ApJS, 201, 23.

\bibitem[Gaskell (1996)]{Gask96} Gaskell, C. M., 1996, \apj, 464, 107.

\bibitem[Gaskell \& Klimek (2003)] {Gask03} Gaskell, C. M. \& Klimek, E. S.,
2003  {Astronomical and Astrophysical Transactions}, 22, 661.
	
\bibitem[Gaskell (2009)] {Gask09} Gaskell, C. M.,2009,  {\nar},  55, 140.


\bibitem[Gillessen (2012)]{cloud} Gillessen, S., 
Genzel, R., Fritz, T. K., Quataert, E., Alig, C., Burkert, A., Cuadra, J.,
Eisenhauer, F., Pfuhl, O., Dodds-Eden, K., Gammie, C. F., Ott, T.
2012, {\nat}, {481}, 51.


\bibitem[Guo et al. (2006)] {Guo06} Guo, D., Tao, J., Qian, B., 2006
{PASJ}, 58, 503.


\bibitem[Hayasaki et al. (2008)]{Hayasaki08}Hayasaki, K., Mineshige, S., Ho, L.
C. 2008, \apj, 682, 1134.


\bibitem[Ho et al. (1995)]{Ho95} Ho, L. C., Filippenko, A. V., Sargent, W. L.,
1995, \apjs, 98, 477.


\bibitem[Katz (1997)]{Katz97} Katz, J.,I., 1997 {\apj}, 478, 527 

\bibitem[Kaspi et al. (1996)] {Kaspi96} Kaspi, S., Maoz, D., Netzer, H.,
Peterson, B. M., Alexander, T., Barth, A. J., Bertram, R., Cheng, F.-Z.,
Chuvaev, K. K., Edelson, R. A., Filippenko, A. V., Hemar, S., Ho, L. C., Kovo,
O., Matheson, T., Pogge, R. W., Qian, B.-C., Smith, S. M., Wagner, R. M., Wu,
H., Xue, S.-J., Zou, Z.-L.,  1996, \apj, 470, 336.


\bibitem[Komossa (2006)]{Komossa}Komossa S., 2006, Mem. Soc. Astron. Ital., 77,
733.

\bibitem[Kraemer et al. (2001)]{krae01} Kraemer, S. B., Crenshaw, D. M.,
Hutchings, J. B.,
    George, I. M., Danks, A. C., Gull, T. R., Kaiser, M. E., Nelson,  
    C. H., Weistrop, D. \& Vieira, G. L., 2001, \apj,
     {551}, 671.

\bibitem[Kraemer et al. (2008)]{krae08} Kraemer, S. B., Schmitt, H. R. \& Crenshaw, D. M.,
      2008, \apj, {679}, 1128.


\bibitem[Lewis et al. (2010)] {lewi10} Lewis, K. T., Eracleous, M. \&
Storchi-Bergmann, T. 2010, {\apjs},  {187}, 416.



\bibitem[Lomb (1976)]{lomb76} Lomb, N. R.,  1976, \apss,
     {39}, 447.


\bibitem[Longo et al. (1996)]{Longo96} Longo, G., Vio, R., Paura, P., Provenzale, A., Rifatto, A., 1996, \aap,{312},424.



\bibitem[Lubinski et al. (2010)] {Lubinski10} P. Lubinski, A. A. Zdziarski, R.
Walter, S. Paltani, V. Beckmann, S. Soldi,
    C. Ferrigno and T. J.-L. Courvoisier, 2010, MNRAS, 408, 1851L.

\bibitem[Lyutyj et al. (1984)]{Lyutyj} Lyutyj, V. M., Oknyanskij, V. L., Chuvaev, K. K.,
1984, {Sov. Astron. Lett.}, {10}, 335 

\bibitem[Malkov et al. (1997)]{Malkov97} Malkov, Yu., F., Pronik, V., I. and 
Sergeev, S., G., 1997, {\aap}, 324, 904 

\bibitem[Marziani et al (2010)] {Marz10} Marziani, P., Sulentic, J. W., Negrete,
C. A., Dultzin, D., Zamfir, S., Bachev, R., 2010, {\mnras}, 409, 1033.

\bibitem[Mayer et al. (2010)] {Mayer10} Mayer, L., Kazantzidis, S., Escala, A.,
Callegari, S., 2010, \nat, 466, 26.

\bibitem[Mayer et al. (2007)] {Mayer07} Mayer, L., Kazantzidis, S., Madau, P., Colpi, M., Quinn, T., Wadsley, J.,
2007, Science, 316, 1874.


\bibitem[Merritt \& Milosavljevi\'c (2005)] {MM05} Merritt, D. \&
Milosavljevi\'c, M., 2005, Living Rev. Relativ., 8, 8.


\bibitem[Milosavljevi\'c \& Merritt (2001)] {MM2001}  Milosavljevi\'c, M. \&
Merritt, D., 2001, {\apj}, {563}, 34


\bibitem[Misner et al. (1973)] {1973grav.book.....M} Misner, C., W., Thorne, K.,
S., \& Wheeler, J., A.\, 1973, Gravitation, {San Francisco: W.H., Freeman and
Co.}, 988.


\bibitem[Mundell et al. (1999)] {mund99} Mundell, C. G., Pedlar, A., Shone, D.
L., Robinson, A. G,  1999, {\mnras}  {304}, 481.

\bibitem[Mundell et al. (2003)] {mund03} Mundell, C. G., Wrobel, J. M., Pedlar,
A. \&   Gallimore, J. F., 2003,      {\apj}  {583}, 192.


    
\bibitem[Newman et al. (1997)] {Newman97}   Newman, J., A.; Eracleous,
M., Filippenko, A., V.; Halpern, J., P., 1997, \apj, 485, 570.

\bibitem[Onken et al. (2007)] {Onken07} Onken, C. A., Valluri, M., Peterson,
B., M., Pogge, R., W., Bentz, M., C., Ferrarese, L., Vestergaard, M., Crenshaw,
D. M,. Sergeev, S., G., McHardy, I., M. et al., 2007, \apj, 670, 105 

\bibitem[Oknyanskij \& Lyuty (2007)] {Okny07} Oknyanskij \& V., Lyuty, V., 2007,
{Perem. Zvezdy Prilozhenie}, 7, 28


\bibitem[Oknyanskij et al (1978)] {Okny78} Oknyanskij, V.L., 1978, {Peremennye Zvezdy}, {21}, 71. 

\bibitem[Pacholczyk et al. (1983)] {Pach83} Pacholczyk, A. G., Penning, W. R.,
Ferguson, D. H.,  Lubart, N. D., Turnshek, D. 1983, Astrophys. Lett., {23},
225.


\bibitem[Penston and Perez (1984)]{Penston84} Penston, M. V. and Perez, E.,
1984,
{\mnras}, {211},  33.

\bibitem[Popovi\'c (2012)] {Pop2012} Popovi\'c, L., \v C., 2012, {\nar}, {56},
74.

\bibitem[Press et al. (1996)] {pres96} Press, W. H., Teukolsky, S. A.,
Vetterling, W. T.  \& Flannery, B. P., 1996,
Numerical Recipes in Fortran 77: The Art of  Scientific Computing. 2nd edn.
Cambridge Univ. Press, Cambridge,
    New York, Melbourne.




\bibitem[Pringle (1996)]{Pring96} Pringle, J., E., 1996, {\mnras}, {281}, 357 


\bibitem[Rainer (1988)]{Velocity} Rainer W.: Velocity 1.3, 1998, A program to
compute radial velocity curves for spectroscopic binary stars, determine
best-fit solutions for measured radial velocity data, and plot the results,
  \emph{http://www.ibiblio.org/pub/Linux/science/astronomy}.



\bibitem[Scargle (1982)]{scar82} Scargle, J. D., 1982, \apj,  {263},
    835.

\bibitem[Sergeev (1994)]{Sergeev94} Sergeev S.G., 1994, Astron. Zh., 71, 189.

\bibitem[Shapovalova et al. (2008)]{shap08} Shapovalova, A.I., Popovi{\'c}, L.{\v C}.,
    Collin, S., {Burenkov}, A.N., {Chavushyan}, V.H., {Bochkarev},
    N.~G. {Ben{\'{\i}}tez}, E., {Dultzin}, D., {Kova{\v c}evi{\'c}},
    A. {Borisov}, N., {Carrasco}, L., {Le{\'o}n-Tavares}, J.
    {Mercado}, A., {Valdes}, J.~R., {Vlasuyk}, V.~V., {Zhdanova},
    V.~E., 2008,   \aap,  {486}, 99.


\bibitem[Shapovalova et al. (2010)]{shap10} Shapovalova, A. I., Popovi\'{c}, L. \v{C}.,
    Burenkov, A. N., Chavushyan, V. H., Ili\'{c}, D.,
    Kova\v{c}evi\'{c}, A., Bochkarev, N. G. \& Le\'{o}n-Tavares, J., 2010, \aap,  {106},  509.


\bibitem[Shen \& Loeb (2010)]{Shen10} Shen, Y. \& Loeb, A. 2010, \apj, {725}, 249S.


\bibitem[Shi et al (2012)]{Shi12} Shi, J., M., Krolik, J., H.,Lubow,
S., H., Hawley, J., F., 2012, \apj, 749, 118.






\bibitem[Storchi-Bergmann et al. (1997)] {Storchi-Bergmann97} Storchi-Bergmann,
T., Eracleous, M., Ruiz, M., T., Livio, M., Wilson, A., S.; Filippenko,
A., V, 1997, \apj, 489, 87.


\bibitem[Sudou et al. (2003)] {Sudou03} Sudou H., Satoru I.,
 Yasuhiro M., Yoshiaki T., 2003,
{Science}, {23}, 1263. 

	
\bibitem[Sulentic et al. (2000)]{Sulentic00} Sulentic, J. W., Zwitter, T., Marziani, P. \& Dultzin-Hacyan, D.
2000, \apj, {536}, 5.



\bibitem[Tsalmantza et al. (2011)] {Ts2011} Tsalmantza, P., Decarli, R., Dotti,
M., Hogg, David W., 2011, {\apj}, {738}, 20 


\bibitem[Ulrich et al. (1985)]{ulri85} Ulrich, M. H., Altamore, A., Boksenberg, A.,
    Bromage, G. E., Clavel, J., Elvius, A., Penston, M V., Perola,
    G. C. \& Snijders, M. A. J. 1985,  {Nature}  {313}, 747.

\bibitem[Ulvestad et al. (1998)]{ulve98} Ulvestad, J. S., Roy, A. L., Colbert,
E. J. M. \&
    Wilson, A. S., 1998, {\apj},  {496}, 196.


\bibitem[Valtonen \& Ciprini (2012)] {Valtonen12} Valtonen, M., Ciprini, S.,
2011, Mem. Soc. Astron. Ital., 83, 219



\bibitem[Zamfir et al. (2010)]{Zamfir10} Zamfir, S., Sulentic, J. W., Marziani, P. \& Dultzin, D.
2010, \mnras, {403}, 1759.







\end{thebibliography}
\end{document}